\documentclass[aps,prl,twocolumn,amsmath,amssymb,amsfonts,reprint,floatfix,superscriptaddress]{revtex4-2}

\usepackage{graphicx}
\usepackage{CJK}
\usepackage{color}
\usepackage[colorlinks,bookmarks=false,citecolor=blue,linkcolor=red,urlcolor=blue]{hyperref}
\usepackage{multirow}
\usepackage{ulem}
\usepackage{tabularx}
\usepackage[nounderscore]{syntax}
\graphicspath{{figs/}}
\newcommand{\ket}[1]{\mbox{$ | #1 \rangle $}}
\newcommand{\bra}[1]{\mbox{$ \langle #1 | $}}
\newcommand{\be}{\begin{equation}}
\newcommand{\ee}{\end{equation}}

\DeclareMathOperator{\Tr}{Tr}

\begin{document}
\title{Sudden change in entanglement Hamiltonian: \\Phase diagram of an Ising entanglement Hamiltonian}
\author{Zhe Wang}
\affiliation{Department of Physics, School of Science and Research Center for Industries of the Future, Westlake University, Hangzhou 310030,  China}
\affiliation{Institute of Natural Sciences, Westlake Institute for Advanced Study, Hangzhou 310024, China}

\author{Siyi Yang}
\affiliation{State Key Laboratory of Surface Physics and Department of Physics, Fudan University, Shanghai 200438, China}
\affiliation{Department of Physics, School of Science and Research Center for Industries of the Future, Westlake University, Hangzhou 310030,  China}
\affiliation{Institute of Natural Sciences, Westlake Institute for Advanced Study, Hangzhou 310024, China}

\author{Bin-Bin Mao}
\affiliation{School of Foundational Education, University of Health and Rehabilitation Sciences, Qingdao 266000, China}

\author{Meng Cheng}
\email{m.cheng@yale.edu}
\affiliation{Department of Physics, Yale University, New Haven, CT 06520-8120, USA}

\author{Zheng Yan}
\email{zhengyan@westlake.edu.cn}
\affiliation{Department of Physics, School of Science and Research Center for Industries of the Future, Westlake University, Hangzhou 310030,  China}
\affiliation{Institute of Natural Sciences, Westlake Institute for Advanced Study, Hangzhou 310024, China}

\begin{abstract}
The form of the entanglement Hamiltonian varies with the parameters of the original system. Whether there is a singularity is the key problem for demonstrating/negating the universality of the relation between the entanglement spectrum and edge energy spectrum. We carefully study the phase diagram of a 1D Ising entanglement Hamiltonian as an example to clarify the long-standing controversy of the general relation between the entanglement Hamiltonian and original Hamiltonian. Interestingly, even if the singularities indeed exist, the Li-Haldane-Poilblanc conjecture, i.e., the general relation between the entanglement spectrum and edge energy spectrum, seemingly still holds.
\end{abstract}

\maketitle

\textit{\color{blue}Introduction.-}
Quantum information science and condensed matter physics have rapidly converged and influenced each other in recent years
~\cite{Amico2008entanglement,Laflorencie2016}. Within this trend, quantum entanglement was employed to reveal fundamental structures, such as field-theoretical and topological properties, of quantum many-body systems~\cite{vidal2003entanglement,Korepin2004universality,Kitaev2006,Levin2006}. For instance, the entanglement entropy (EE) has become an important tool to characterize topological states and conformal field theories (CFT)~\cite{Calabrese2008entangle,Fradkin2006entangle,Nussinov2006,Nussinov2009,CASINI2007,JiPRR2019,ji2019categorical,kong2020algebraic,XCWu2020,JRZhao2020,XCWu2021,song2023extracting,song2023deconfined,deng2024diagnosing,d2024entanglement,song2024quantum}. 
Beyond the EE, Li and Haldane proposed that the entanglement spectrum (ES) is a more organic entanglement characteristic to study the intrinsic information of many-body systems~\cite{Li2008entangle,Thomale2010nonlocal,Poilblanc2010entanglement}. 
Thereafter, low-lying ES has been widely used as a key signature to investigate CFT and topological order in highly entangled quantum matter~\cite{Pollmann2010entangle,Fidkowski2010,Yao2010,XLQi2012,Canovi2014,LuitzPRL2014,LuitzPRB2014,LuitzIOP2014,Chung2014,Pichler2016,Cirac2011,Stoj2020,guo2021entanglement,Grover2013,Assaad2014,Assaad2015,Parisen2018,yu2021conformal,RoosePrb2021,RoosePrb2022}.

Going beyond EE and ES, the concept of entanglement Hamiltonian (EH) was also proposed to better describe the structure of bipartite entanglement. Its definition is $H_E=-\ln(\rho_A)$, where the $\rho_A$ is the reduced density matrix (RDM). Both EE and ES can be understood as some observables of the EH at certain temperature. In this sense, the EH can be treated as an even more basic physical quantity and is highly worthy to be carefully studied.
Recently, further detailed studies have analyzed the EH based on field theory, numerics, and experiment~\cite{alba2012entanglement,zhu2020entanglement,zhu2019reconstructing,Tang2020critical,Mendes_Santos_2020,Dalmonte_2022,Joshi2023,redon2023realizing,dalmonte2018quantum,ma2023multipartite,Eisler_2020,PhysRevB.100.195109,li2023relevant,mao2023sampling,song2023different,zyan2021entanglement,wu2023classical}.

The correspondence between the EH and the original Hamiltonian (OH) is still a charming puzzle that hasn't been completely solved yet. The famous Li-Haldane-Poilblanc (LHP) conjecture~\cite{Li2008entangle,Poilblanc2010entanglement} posited that the EH always resembles the OH on the virtual edge. Furthermore, the Bisognano-Wichmann (BW) theorem gives the form of the EH based on its OH, but it only holds under some strict conditions, e.g., the entanglement region has no corner, the low-energy theory is a relativistic quantum field theory~\cite{dalmonte2022entanglement,mendes2019entanglement,Mendes-Santos_2020,giudici2018entanglement}. In both cases (i.e. when the LHP conjecture holds, or when the BW theorem applies), the EH should change smoothly when continuously tuning the parameter of the original Hamiltonian.

However, Chandran, Khemani and Sondhi argued that the EH could have singularities at quantum critical points (QCP) of the original Hamiltonian~\cite{chandran2014how}. A straightforward example they proposed is a discussion about the EH of a 2D transverse field Ising model (TFIM) with half-cutting. In their argument, according to the Mermin-Wagner (MW) theorem~\cite{mermin1966absence,mermin1967absence}, the system of EH can not be ordered except at $T_E=0$ (temperature for the EH) in the paramagnetic (PM) phase of the OH because the EH of a 2D TFIM is a 1D Ising-like system~\footnote{According to the Mermin-Wagner (MW) theorem, a 1D Ising chain with short range interaction can not be ordered in finite temperature}. 
However, at $T_E \geq 1$, the system of the EH presents an Ising order in the ordered phase of the OH (i.e., the TFIM). Notice that $\rho_A=e^{-H_E}$ can be treated as the ensemble of $H_E$ at $T_E=1$. Meanwhile, the physical property of this ensemble should be same as the ground state of the original Hamiltonian. 

In the above analysis, the system of the EH in PM side of the OH must be disordered when $T_E>0$ and is ordered in the Ising side in the region $T_E\leq 1$, thus the EH here should display a discontinuity at the QCP of the original Hamiltonian. While physically reasonable, the argument relies on approximations that are not fully justified.
 For example, the EH was derived in the strong coupling limit by first-order perturbation theory,  which is shown to be short range. However, near the QCP (corresponding to the thermal phase transition of the EH), the long-range part of the interactions of the EH should be significantly enhanced~\footnote{Only in the presence of long-range interactions, the 1D Ising system can violate MW theorem and has a phase transition.}, which are not captured by the perturbative analysis.

Meanwhile, the importance of long-range interactions was highlighted by a recent numerical work~\cite{li2023relevant}. It was shown that, the long-range terms of the EH can indeed change the underlying physics, $e.g.$, violating the Mermin-Wagner theorem. 
This motivates a thorough investigation of long-range interactions and the corresponding phase diagram of the EH within a concrete lattice model.
In this paper, we take up this challenge by studying a quantum Ising model through both numerical calculations and theoretical analysis. Our results provide new insights into the nature of the ES and confirm that the universal features of the ES remain robust, even under these modifications.

\begin{figure}[htb]
	\centering
	\includegraphics[width=0.46\textwidth]{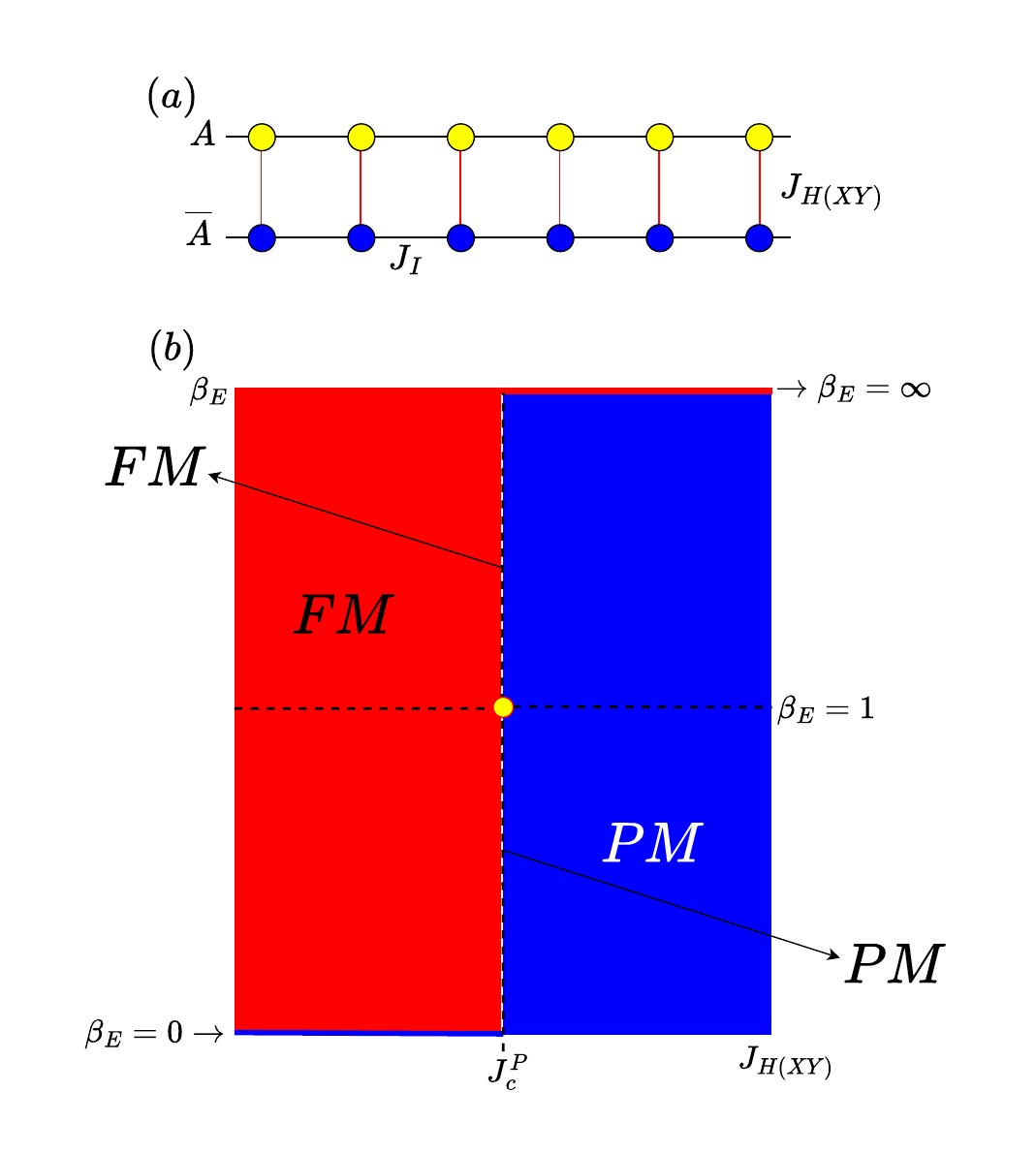}
	\caption{(a) Ising ladder: the chains with FM Ising exchange interactions $J_{I}$ coupled by  AFM rung Heisenberg couplings $J_{H}$ or XY-like couplings $J_{XY}$. We choose the chain with blue sites as environment $\overline{A}$ and the chain with yellow sites as the subsystem $A$ we concern. (b) Phase diagram of the entanglement
Hamiltonian $H_{E}$  of the original Hamiltonian $H_H(H_{XY})$.  The yellow  dot at ($J_{c}^P$, $\beta_E=1$) represents a quantum transition of the original Hamiltonian   or  thermodynamic phase transitions of the EH. The red region indicates that the system is in the FM phase, and the blue region indicates the paramagnetic (PM) phase. Two long arrows indicate that the dashed line belongs to the FM or PM phase, respectively.}
	\label{Fig:modelanddiagram} 
 \end{figure}

\textit{\color{blue} The models and methods.-} The system we discussed here is a spin-1/2 ladder [Fig. \ref{Fig:modelanddiagram}(a)], with intra-chain ferromagnetic (FM) Ising interactions, and inter-chain antiferromagnetic (AFM) Heisenberg interactions. The Hamiltonian is given by
\begin{equation}
	H_H = J_H \sum_i {S}_{i,1}\cdot {S}_{i,2}
	+J_I \sum_{l=1,2} \sum_i S^y_{i,l}\cdot S^y_{i+1,l}
		\label{H_ladder}
	\end{equation}
where $l=1,2$ denote the two legs of the ladder, $J_H>0$ is the inter-chain AFM Heisenberg exchange interaction, and $J_I<0$ is the intra-chain FM Ising interaction. This model can be efficiently simulated by a cluster Stochastic Series Expansion (SSE) algorithm~\cite{wu2023classical,sandvik2010computational,yan2019sweeping,yan2020improved}.



We will also consider a variant of Eq.(\ref{H_ladder}), where the inter-chain interactions are XY-like:
\begin{equation}
	\begin{split}
	H_{\rm XY} &= -J_{XY} \sum_i ({S}^x_{i,1}\cdot {S}^x_{i,2}+{S}^y_{i,1}\cdot {S}^y_{i,2})\\
	&+J_I \sum_{l=1,2} \sum_i S^y_{i,l}\cdot S^y_{i+1,l}
		\end{split}
		\label{H_ladder2}
\end{equation}
 The advantage of this model is that the reduced density matrix can be found exactly, thus facilitating a more systematic study of the phase diagram of the EH. The details of the derivation of the reduced density matrix is given in the Supplementary Materials (SM). It is worth noting that both the two models above have similar Ising-like EHs and related physics.

We choose one chain as the environment and the other one as the subsystem to further explore the entanglement property. Via an analysis of the path-integral representation for this class of models~\cite{wu2023classical}, the reduced density matrix must be diagonal in the ${S}^z$ basis, thus the EH is effectively a classical Ising model (with possibly long-range interactions). This is shown explicitly for the second model $H_{\rm XY}$ (see SM for details).

For $H_{\rm H}$ in Eq.(\ref{H_ladder}), we use the recently developed replica quantum Monte Carlo method~\cite{li2023relevant,zyan2021entanglement,song2023different,wu2023classical,liu2023probing} to study the ensemble of EH. The key point is to simulate a R\'enyi replica partition-function $Z_A^{(n)}\propto \Tr \rho_A^{n}=\Tr e^{-n H_E}$. The replica index $n$ here can be considered as the effective inverse temperature $\beta_E=1/T_E$ of the EH $H_E$. By simulating different values $n$, one can effectively tune the temperature of the EH. More details of this method are explained in the SM. Note that in this article, when we talk about $\beta_E$, we mean the inverse temperature of EH, and $\beta$ is the inverse temperature of OH. 

Although the replica QMC algorithm can perform large-size simulations, the effective inverse temperature $\beta_E$ can only take integer values and $\beta_E \geq 1$, which limits our ability to explore the phase diagram of the entanglement Hamiltonian at high temperature. 
To this end, the other model Eq.(\ref{H_ladder2}),  provides a complementary perspective: because the RDM is explicitly known, the EH at any temperature can be studied. However, the high computational complexity of calculating the weights in the reduced density matrix makes that the available system size is smaller than using the replica QMC method. By combining these two methods, we can synthetically explore the phase diagram of the Ising-like entanglement Hamiltonian.

\textit{\color{blue}Numerical results.-} The phase diagram of the EH can be studied as a function of the inverse temperature $\beta_E$ of the EH and the coupling $J_H$($J_{\rm XY})$ of the OH. First, we note that the two models show qualitatively the same phase diagram, which consists of ferromagnetic (FM) and paramagnetic (PM) phases.
The Ising magnetic order can be characterized by the uniform magnetization $ m^y=\frac{1}{L}\sum_i  S_i^y$ (the FM ordered along the $y$ axis in our models). We define the Ising Binder cumulant $U_{2}= \frac{3}{2}\left(1-\frac{1}{3}\frac{\langle (m^{y})^{4}\rangle}{\langle (m^{y})^{2}\rangle^{2}}\right)$. 
It is well-known that in the FM (PM) phase, $U_{2}$ converges to 1 (0) with increasing system size. The critical point can be accurately located by analyzing the crossing of $U_2$. 

\begin{figure}[htb]
	\centering
	\includegraphics[width=0.5\textwidth]{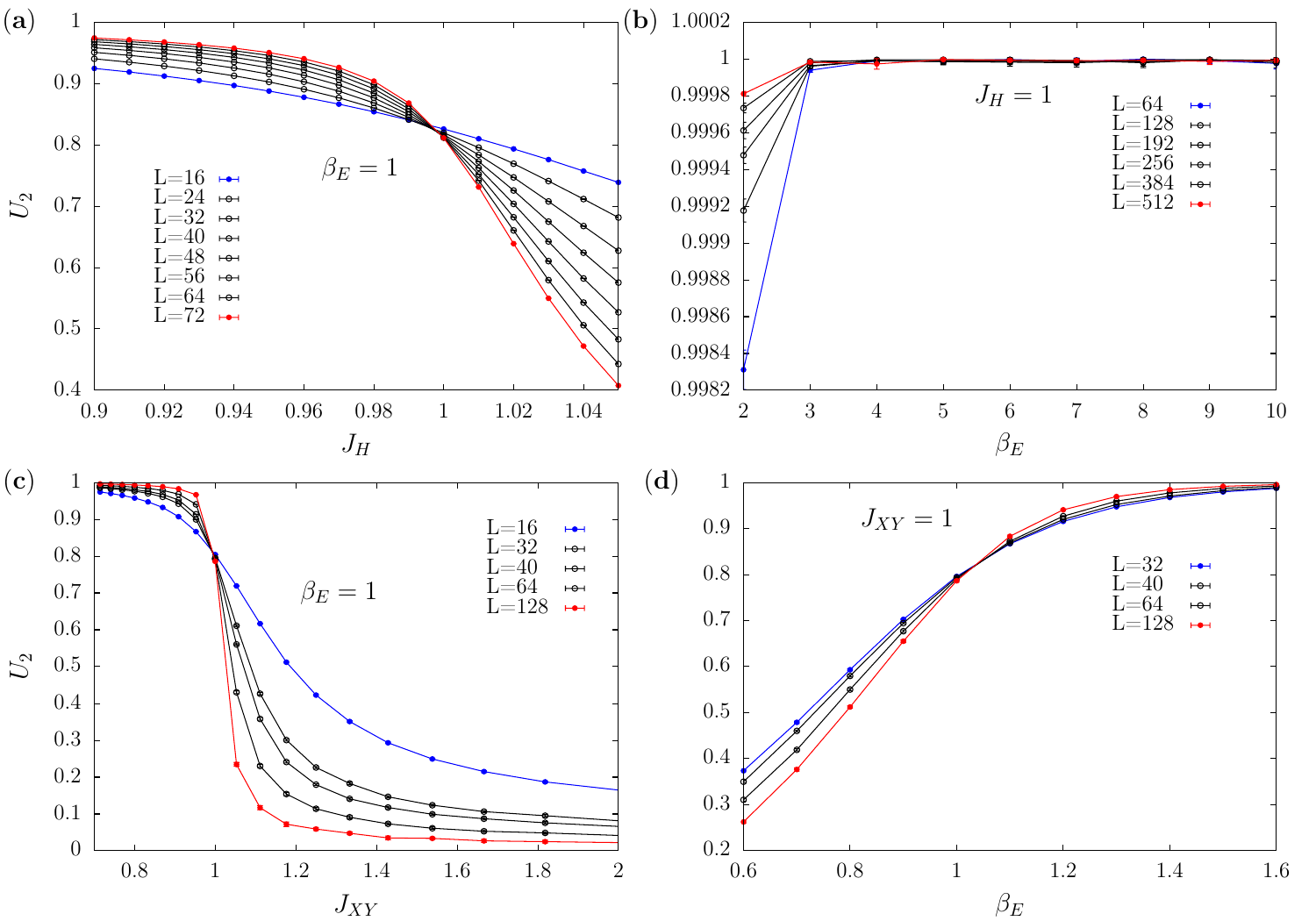}
	\caption{The Binder cumulant are calculated along two dashed lines passing through QCPs of OH in Fig. \ref{Fig:modelanddiagram}(b). (a) and (b) are the results of the EH from Hamiltonian $H_H$. (c) and (d) are the results of the EH from Hamiltonian $H_{XY}$.}
	\label{Fig:qcp} 
 \end{figure}

Let us start from the the line of $\beta_E=1$ in the Fig.\ref{Fig:modelanddiagram} (b). By definition, the phase diagram should be same as the one of the original Hamiltonian. Denote by $J_c^P$ the critical coupling of the OH, the right side of which is the PM phase and the left is FM. By setting $J_I=-1$ in  Eq.(\ref{H_ladder}) and $J_I=-1/2$ in Eq.(\ref{H_ladder2}), the QCPs of the two models should have same value $J_c^P=1$. This is confirmed numerically in both models. For $H_{\rm H}$, the QMC calculation of $U_2$ in the ground state is shown in [Fig. \ref{Fig:qcp} (a)]. For $H_{\rm XY}$, results from the classical MC simulation of the EH are presented in Fig. \ref{Fig:qcp} (c).

At $\beta_E=1$, the quantum critical point of the OH is equivalent to the thermal critical point of the EH at $J_c^P$. The existence of such a thermal phase transition implies that the EH at $J_c^P$ must have long range interactions in order to circumvent the Mermin-Wagner theorem. This is further demonstrated directly in explicit form of the EH given by Eq.(\ref{H_ladder2}) in the SM.

Having understood the phase diagram along $\beta_E=1$, we now explore the global phase diagram more systematically.  First we consider the neighbohood of the critical point at $\beta_E=1, J_H(J_{XY})=J_c^P$. An immediate question arises: away from $\beta_E=1$, is $J_H(J_{XY})=J_c^P$ an isolated critical point, or it belongs to a critical line?

To answer the above question, we fix the $J_H (J_{XY}) =J_c^P=1$ and tune $\beta_E$ to study the behavior of the Binder cumulant, as displayed in Fig. \ref{Fig:qcp} (b) and (d). Although the replica method of QMC can only access integer values of $\beta_E \geq 1$, from Fig. \ref{Fig:qcp} (b) we can clearly see that the system (EH of Eq.(\ref{H_ladder})) is ordered when  $\beta_E >1$. For $H_{\rm XY}$, the analytical form of the EH (Eq.(\ref{H_ladder2})) allows simulations done in whole parameter region, which shows clearly that the thermal phase transition only happens at $\beta_E=1$ [Fig. \ref{Fig:qcp} (d)]. One can find that the physics shown in (b) is qualitatively similar to the one in the region $\beta_E>1$ of (d). Both figures support the conclusion that there is a single critical point along the $J_c^P$ line instead of a critical line.


Furthermore, we adopt finite-size scaling $U_2(g,L)=L^{-\kappa}f[(g-g_{ c})L^{1/\nu}]$  to  rescale the Binder cumulant. Here $g=J_H (J_{XY})$ or $T_E=1/\beta_E$, and $\nu$ is correlation length exponent. Using the 2D Ising critical exponent $\nu=1$, we find that the data (obtained by setting $\beta_E=1$ and tuning $J_H(J_{XY})$, see horizontal dashed line in Fig. \ref{Fig:modelanddiagram} (b)) of $U_2$ for different  system sizes fall onto the same curve (see SM), which means that the critical point on the $\beta_E=1$ line belongs to the 2D Ising universality class, as expected.

However, the critical exponents become totally different if the critical point is approached along another direction. Setting $J_{XY}=1$ and tuning $\beta_E$ (the vertical dashed line at $J^P_c$ in Fig. \ref{Fig:modelanddiagram} (b)), $U_2$ for different  system sizes can not be collapsed to a same curve with $\nu=1$. If we set $\nu$ free to fit the data, the optimal value is $1/\nu=0.24(1)$. Additionally, we perform scaling collapse for ${\langle (m^{z})^{2}\rangle}$ with $\kappa=2\beta/\nu$ to fit $\nu$ and $\beta/\nu$. We find $\beta/\nu=0.123(2)$ and  $1/\nu=0.34(2)$. The former is consistent with 2D Ising universality class, while the latter is not. Note that the $1/\nu$ fitted from $(m^z)^2$ is also different from $1/\nu=0.24(1)$ from the scaling analysis of $U_2$ above, which may result from finite-size effects. 
Details about the fitting of critical exponents can be found in SM. The fitting results show $\beta/\nu$ is universal in different parameter paths ($\beta_E$ or $J$), while $\nu$ is highly dependent on the detailed path. The reason may be either the critical exponents $\beta$ or $\nu$ explicitly includes the effective dimension $d$ of the Hamiltonian, but $\beta/\nu$ is dimensionless. 

\begin{figure}[htb]
	\centering
	\includegraphics[width=0.5\textwidth]{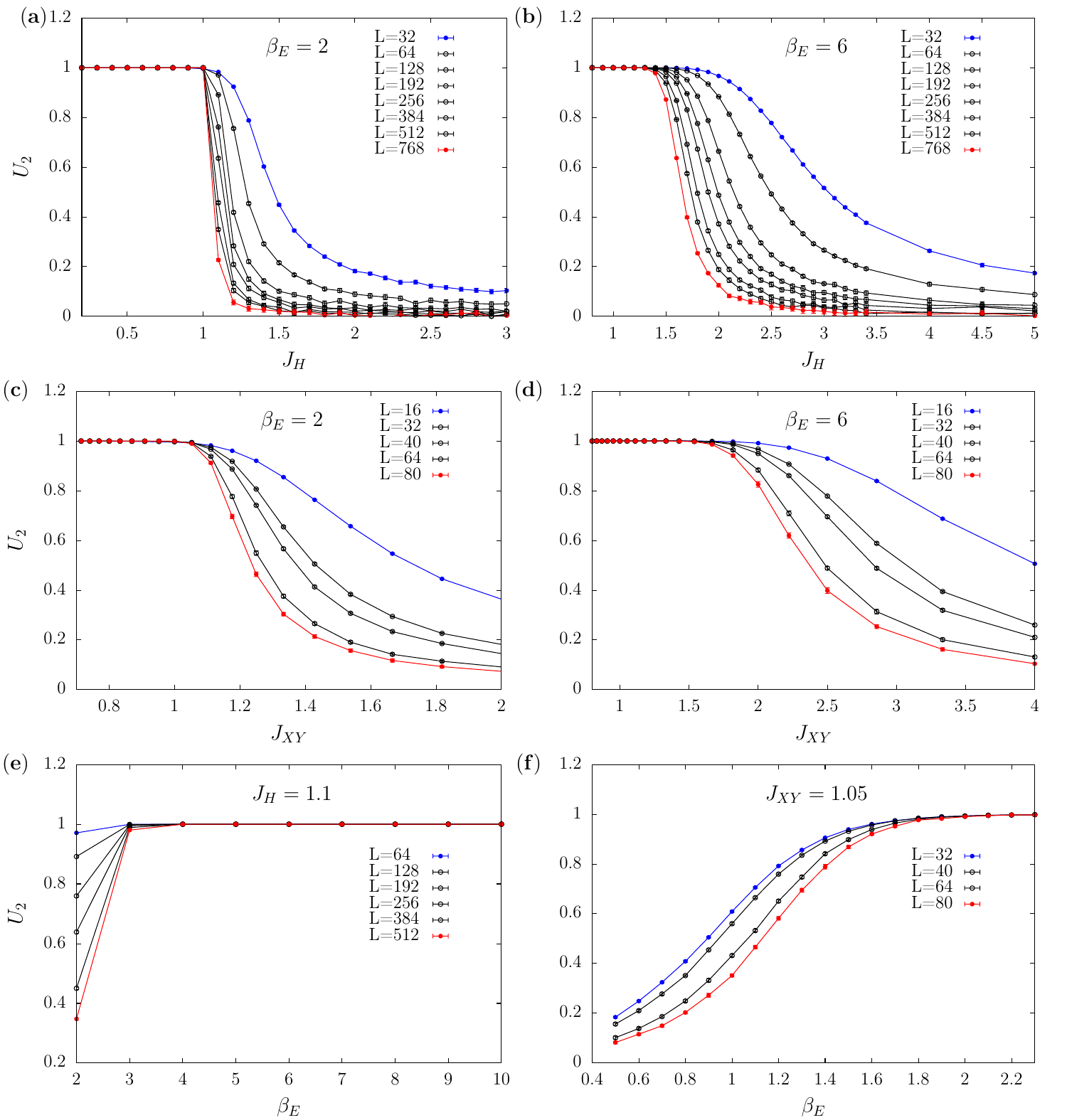}
	\caption{  The Binder cumulant $U_{2}$  results  of the EH for different $L$ (a) under $\beta_E=2$  sweeping $J_H$ and (b) under $\beta_E=6$  sweeping $J_H$ and (c) under $\beta_E=2$  sweeping $J_{XY}$ and (d) under $\beta_E=6$  sweeping $J_{XY}$ and (e)  under $J_{H}$=1.1  sweeping $\beta_E$ and (f) under $J_{XY}$=1.05  sweeping $\beta_E$.  Sweeping $J_H(J_{XY})$ or $\beta_E$, there are no crossings of curves with different $L$ for both systems, which like the 1D ising model at the finite temperature (see SM). This means no phase transition when focus on OH in the PM phase.     }
	\label{Fig:pm} 
 \end{figure}

Now the question becomes whether the EH will turn into a short-range one suddenly when it is moved  slightly to the right side of $J_c^P$. To understand this issue, we fix $\beta_E$ and scan the parameter $J_H(J_{XY})$ around $J_c^P$ to obtain the Binder cumulant $U_2$, as shown in Fig. \ref{Fig:pm}. Via the replica method, the simulated sizes are much larger. Fig. \ref{Fig:pm} (a) and (b) show that the Binder cumulant has no crossing point and the curve becomes sharper and sharper when the system size increases. Similar results have been shown in the figure (c) and (d) through simulating the analytical EH of OH $H_{XY}$, although the simulated system sizes are smaller in this approach. Then we choose the parameter points close to $J_c^P$ in both two models, $J_H=1.1$ and $J_{XY}=1.05$, to simulate the $U_2$ in the respective methods. Either figure (e) or (f) has no obvious crossing, which suggests that there is no phase transition at finite temperature. This is consistent with the expectation from the Mermin-Wagner theorem in systems with short-range interactions.
Therefore, our simulation results support the conclusion that there is a sudden change of the EH before the $J_c^P$, that is, the EH remains short-range until reaching the QCP of the OH. 

The numerical results can also be understood theoretically. As the Fig. \ref{Fig:gxjxy} 
in the SM show, the form of the analytical EH reflects that all the matrix elements $g(x,L)$ are flat except near edges $x=0$ or $L$. The physical meaning of a element $g(x,L)$ is the energy cost between two domain walls with distance $x$. The flat curve actually represents there is no energy cost when moving a domain wall (if we ignore the edge effect). It is an obvious short-range effect in a 1D Ising chain, which strongly supports the EH here is short-range. 

\begin{figure}[htb]
	\centering
	\includegraphics[width=0.5\textwidth]{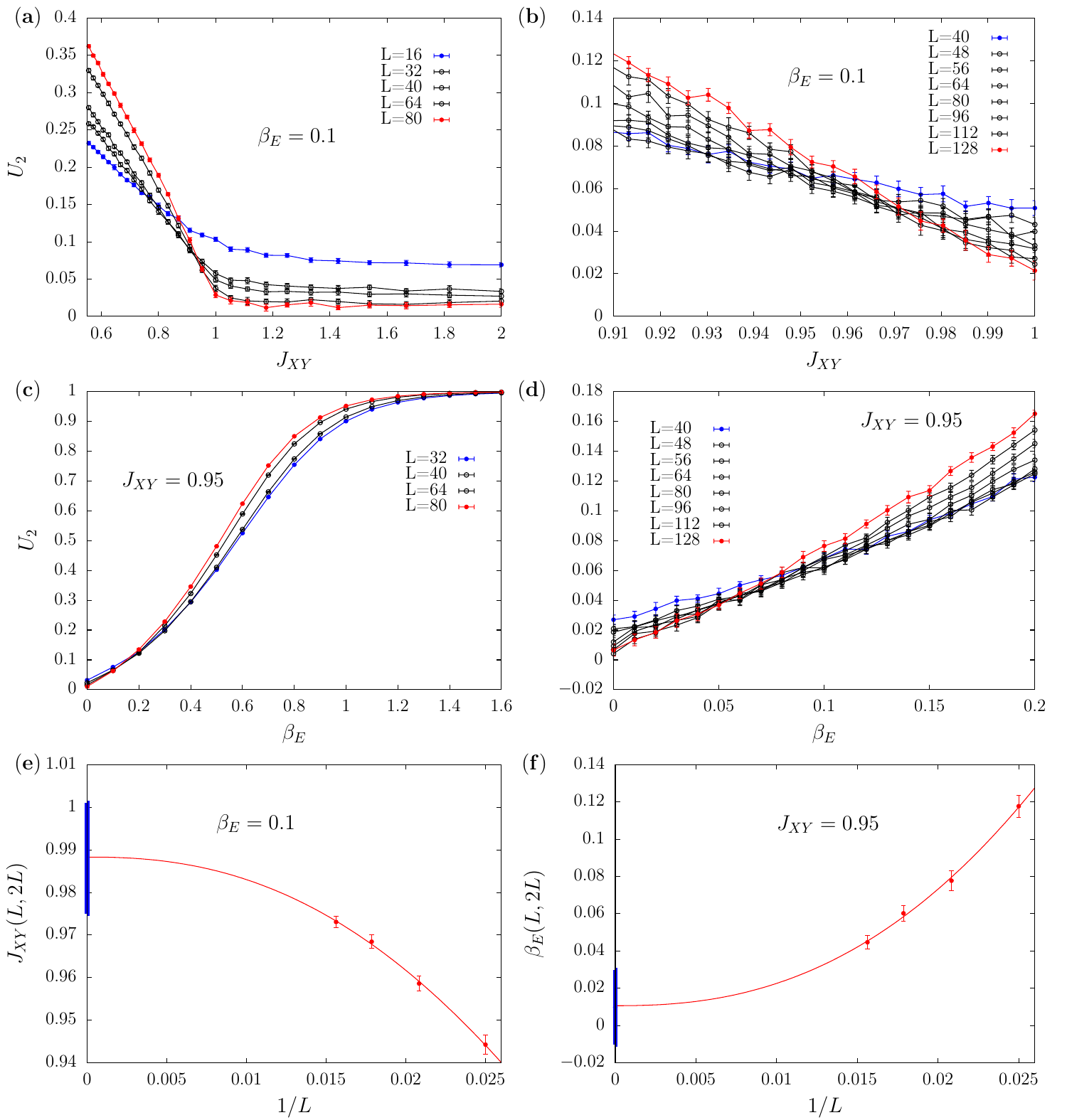}
	\caption{ We focus on the OH of the  system $H_{XY}$ in the FM phase. The Binder cumulant $U_{2}$  results  of the EH for different $L$ (a) under $\beta_E=0.1$  sweeping $J_{XY}$ and (b) shows more detailed
data about (a) and (c) under $J_{XY}=0.95$ sweeping $\beta_E$ and (d) shows more detailed
data about (c). (e) Fitting about  the position of the  crossing points  of the $U_2$  curves setting $\beta_E=0.1$. The blue symbol shows the extrapolated value at the
thermodynamic limit, which is 1 within the error bar. (f) Fitting about the position of the  crossing points  of the $U_2$  curves setting $J_{XY}=0.95$.  The blue symbol shows the extrapolated value at the
thermodynamic limit, which is 0 within the error bar. Details about the  fitting see SM.} 
	\label{Fig:fm} 
 \end{figure}

Then the next question is whether the finite temperature phase transition boundary of the FM and PM phases on the left side is continuously connected to the critical point at $J_c^p$ and $\beta_E=1$ or not? In this case, we need focus on the high temperature region $\beta_E<1$ where the replica method is powerless, thus we use the classical Monte Carlo to simulate the second EH, Eq.(\ref{H_ladder2}), exactly obtained from theoretical analysis. Firstly, we fix the $J_{XY}=0.95$ and scan the $\beta_{E}$ to explore the Binder cumulant crosses, as shown in Fig. \ref{Fig:fm} (c) and (d). Although the crossing behaviour was found weakly, the shift of the crossing point of different system sizes are obvious. According to the data extrapolation in the Fig. \ref{Fig:fm} (f) and SM, the crossing point moves to the $\beta_E=0$ as $L\to\infty$, which reveals the FM phase is robust in any finite temperature. We then fix the $\beta_{E}=0.1$ and scan the $J_{XY}$ (see Fig. \ref{Fig:fm} (a) and (b)). Similarly, we can clearly see the drift of the  crossing point of different system sizes. According to the data extrapolation in the Fig. \ref{Fig:fm} (e) and SM, the crossing point  tends to the $J_{XY}=1$. As mentioned above, the EH is in disorder phase at the ($J_{XY}=1$, $\beta_{E}=0.1$), which means there is no phase transition. 

Therefore, as shown in Fig. \ref{Fig:modelanddiagram}, the PM only exists at exactly $\beta_E=0$ in the region $J<J_c^P$ and FM only lives at strictly $\beta_E=\infty$ in the region $J>J_c^P$.

\textit{\color{blue} Discussion and Conclusion.-}
To summarize, we have systematically studied the full phase diagram of the EH (Ising-like Hamiltonian) obtained from the ground state of the quantum Ising ladder systems. Although for a specific model system, our results shine new light on the relation between the EH and OH. 
For example, we now understand that the EH can exhibit abrupt changes even when the OH parameters are varied smoothly. Additionally, long-range interactions can arise suddenly  in the EH under certain conditions.
 

Nevertheless, we observe that the phase diagram is still qualitatively consistent with the LHP conjecture. For example,
 at the zero temperature $\beta_E =\infty$, the ground state of the EH is always FM, no matter how large $J_{H}(J_{XY})$ is. This suggests the low-lying excitations of the ES should be similar in the whole parameter region. The LHP conjecture predicts that the ES resembles the low-energy spectrum of Hamiltonian on the virtual edges. In our case, the Hamiltonian of the virtual edge is a single FM Ising chain, thus the low-lying ES always displays the excitation of FM state no matter how the inter-chain coupling changes.

In summary, we have studied the phase diagram of an entanglement Hamiltonian in detail for the first time via solid numerical simulation combined with exactly theoretical solution. The sudden changes of the EH along the parameter $J_{H}(J_{XY})$ of the OH, emergent long-range interactions in the EH and unusual critical exponents reveal the rich physics in the world of entanglement Hamiltonian. 
These findings provide deeper insights into the structure of the EH and open new avenues for its exploration.

\textit{\color{blue}Acknowledgement.-}
We acknowledge the start-up funding of Westlake University. The authors also acknowledge Beijing PARATERA Tech Co.,Ltd.(\url{https://www.paratera.com/}) for providing HPC resources that have contributed to the research results reported within this paper. Z.W. is supported by the China Postdoctoral Science Foundation under Grants No.2024M752898.

\clearpage
\appendix
\setcounter{equation}{0}
\setcounter{figure}{0}
\renewcommand{\theequation}{S\arabic{equation}}
\renewcommand{\thefigure}{S\arabic{figure}}
\setcounter{page}{1}
\begin{widetext}
\linespread{1.05}
	
\centerline{\bf\Large Supplemental Material} 

\section{Derivation of the entanglement Hamiltonian}

Consider the following model(for convenience we set $J_I=-\frac12$):
\begin{equation}
    H = -\frac12\sum_{l=1,2}\sum_{j}S^y_{j,l}S^y_{j+1,l}-J_{XY}\sum_j (S^x_{j,1}S^x_{j,2}+S^y_{j,1}S^y_{j,2}).  
\end{equation}

Notice that $[H,S^y_{j1}S^y_{j2}]=0$ for all $j$, so the ground state should reside in the subspace with $S^y_{j1}=S^y_{j2}$. Define 
\begin{equation}
    \tilde{X}_j = S^x_{j,1}S^x_{j,2}, \tilde{Z}_j=S^y_{j,1},
\end{equation}

we find within the subspace (or similarly in any such subspace)
\begin{equation}
    H = - \sum_j (\tilde{Z}_j\tilde{Z}_{j+1}+J_{XY}\tilde{X}_j) - J_{XY}L.
\end{equation}
This is nothing but the transverse-field Ising (TFI) model. Suppose the ground state wavefunction of the TFI model in the $\tilde{Z}$ basis is $\psi(s)$, where $s$ is a bit string. Then the ground state of $H$ is given by 
\begin{equation}
    \ket{\psi}=\sum_s \psi(s) |s\rangle_1|s\rangle_2.
\end{equation}
Tracing out one chain, e.g. the 2nd chain, we find the reduced density matrix of the first chain being 
\begin{equation}
\label{eq7}
    \rho_A = \sum_s |\psi(s)|^2 \ket{s}\bra{s}.
\end{equation}
Clearly, the entanglement Hamiltonian is fully classical. 

To compute $|\psi(s)|^2$, we note that it must satisfy $\psi(s)=\psi(-s)$ due to the Ising symmetry. As a result, $\psi(s)$ only depends on the configuration of domain walls $\tau_j = s_js_{j+1}$. Using the Kramers-Wannier duality:
\begin{equation}
    \tau^x_j=\tilde{Z}_j\tilde{Z}_{j+1}, \tau^z_{j-1}\tau^z_{j}=\tilde{X}_j,
\end{equation}
Note that with PBC we must have $\prod_j \tilde{X}_j=1$ and $\prod_j\tau^x_j=1$. It is clear that the (finite-size) ground state of the transverse-field Ising chain always has $\prod_j\tilde{X}_j=1$.

\begin{figure}[htb]
	\centering
	\includegraphics[width=0.6\textwidth]{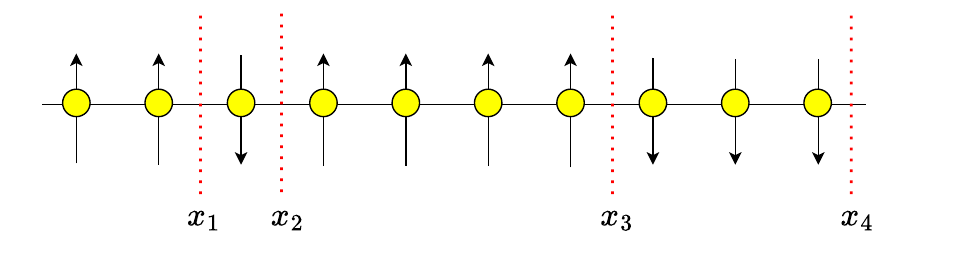}
	\caption{ A typical configuration of the ten-spin chain is used to explain the position of it's domain wall $x_1$,$x_2$...$x_N$. We label the $x_N$ by the spin in front of the domain wall.}
	\label{Fig:domain} 
 \end{figure}
 
After duality transformation the Hamiltonian becomes
\begin{equation}
    \frac{1}{J_{XY}}H = -\sum_j(\tau^z_{j}\tau^z_{j+1}+ \frac{1}{J_{XY}}\tau^x_j).
\end{equation}
Therefore 
\begin{equation}
    \psi(s)\equiv \psi(\tau), 
    \label{eq10}
\end{equation}
where $\tau$ is a bit string of $\tau^x$. We can then rewrite the Hamiltonian in terms of Jordan-Wigner fermions:
\begin{equation}
    \frac{1}{J_{XY}}H=-\sum_j (c_j^\dag-c_j)(c_{j+1}+c_{j+1}^\dag)-\frac{1}{J_{XY}}\sum_j (1-2c_j^\dag c_j).
\end{equation}
Here $\tau^x=1-2n_j$. When $J_{XY}\rightarrow \infty$, this is the Majorana chain Hamiltonian, corresponding to the ferromagnetic state of the dual TFI model.

The ground state of the fermion model takes the following form:
\begin{equation}\label{eq12}
    \exp \left(\frac12 \sum_{ij}g_{ij}c_i^\dag c_j^\dag \right)\ket{0},
\end{equation}
where $\ket{0}$ is the empty state with $n_j=0$ for all $j$, or $\tau_j^x=1$. $g_{ij}$ can be understood as the wavefunction of a Cooper pair. If there are $N$ fermions at $x_1,x_2, \cdots, x_N$, then
\begin{equation}
    \psi(x_1,x_2,\cdots, x_N)=\mathrm{Pf}\, g_{x_i,x_j}.
\end{equation}
In the direct product state representation of spins,   $x_1$,$x_2$...$x_N$ is the position of its domain wall (example see Fig. \ref{Fig:domain}).

According to Eq.(\ref{eq10}) and (\ref{eq12}), $|\psi(s)|^2$ in  Eq.(\ref{eq7}) is the value of the determinant  as follows
\begin {scriptsize}
\begin{equation}
\left | \begin{matrix}
0 & g(x_1-x_2) & \cdots & g(x_1-x_N)\\
g(x_2-x_1) & 0 &\cdots & g(x_2-x_N)\\
\vdots&\vdots & \ddots &\vdots \\
g(x_N-x_1) & \cdots & g(x_N-x_{N-1}) & 0 \\
\end{matrix} \right |
\end{equation}
\end {scriptsize}
where $g(x) = \frac{2}{L}\sum_{n=1}^{L/2} \bigg[ \sin(k x)
	\sqrt{\frac{E(k)-\xi(k)}{E(k)+\xi(k)} } \bigg]$ in which $L$ is the system size, $k=(2n-1)\pi/L$,  $\xi(k)=-2\cos(k)+2/J_{XY}$ and  $E(k)=\sqrt{\xi^2(k)+4\Delta^2 \sin^2k}$. 

\begin{figure}[htb]
	\centering
	\includegraphics[width=0.46\textwidth]{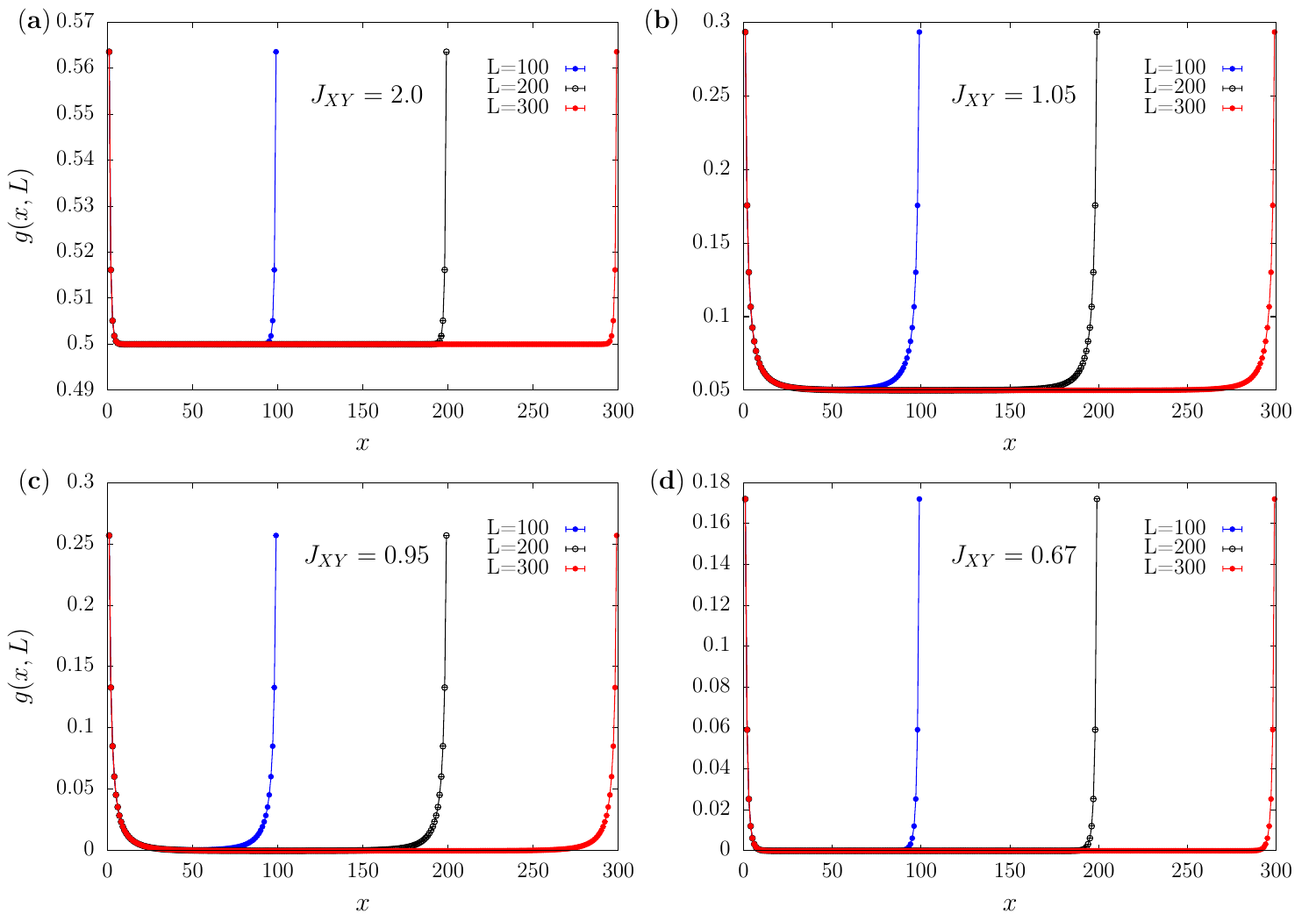}
	\caption{ Matrix element g(x,L) in the PM (a)(b)  or FM (c)(d) phase of the OH.} 
	\label{Fig:gxjxy} 
 \end{figure}
 
 Since  $\rho_A=e^{-H_E}$, $\bra{s} e^{-H_E}\ket{s}=|\psi(s)|^2$ according to Eq.(\ref{eq7}). For arbitrary  effective inverse temperature $\beta_E$ of the EH $H_E$,  we can derive the weight $W_s$ of each  $\left \vert s \right \rangle$ as $\bra{s} e^{- \beta_E H_E}\ket{s}=|\psi(s)|^{2\beta_{E}}$. Using this weight, we can simulate the   EH by  MC method.

Notice that the physical meaning of the element $g(x)$ is similar to the energy cost between two domain walls with distance $x$ on the Ising chain. As drawn in Fig. \ref{Fig:gxjxy} (a) and (b), the $g(x)$ of different system size $L$, i.e., $g(x,L)$, has been plotted. As the size $L\rightarrow \infty$, ignoring the edge effect, the $g(x)$ in the bulk remains a finite value. It actually points to a long-range interaction of the Ising chain. Conversely, Fig. \ref{Fig:gxjxy} (c) and (d) show the $g(x)$ in the bulk approaches to 0 which means a short-range interaction of the EH.


 \section{The universality class at the critical points }
 \begin{figure}[htb]
	\centering
	\includegraphics[width=0.46\textwidth]{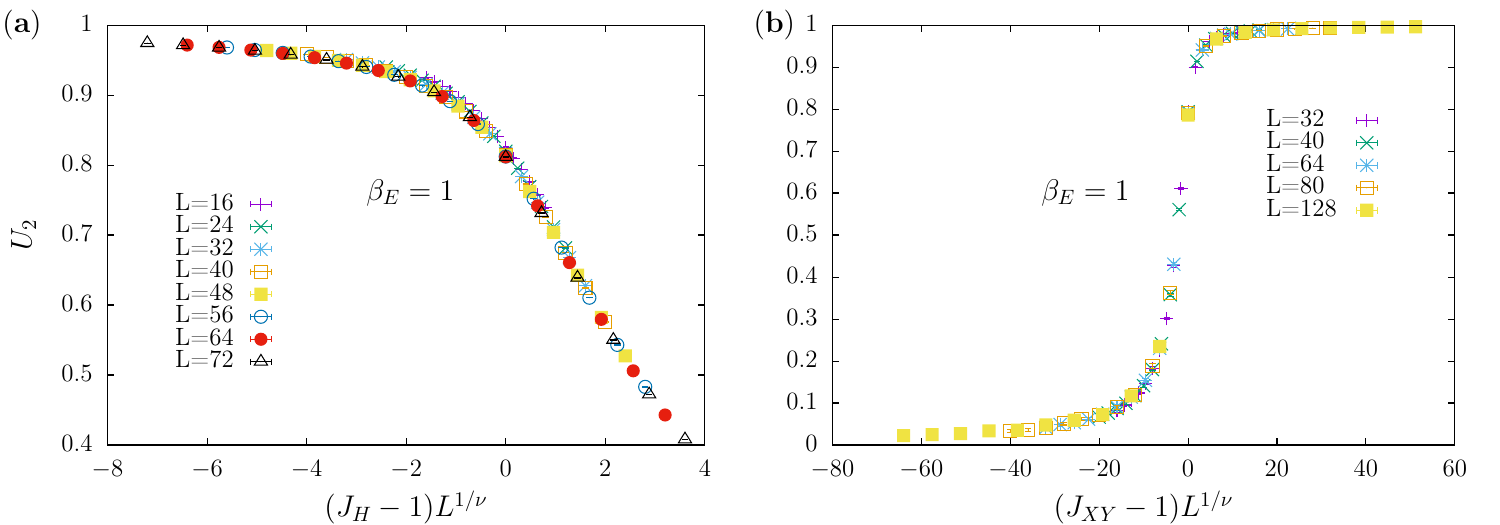}
	\caption{Setting $\beta_E=1$ and sweeping $J_{H}(J_{XY})$ near the critical point $J_{c}^P=1$, the data  of $U_2$ for different  system sizes fall onto the same curve with $\nu=1$. } 
	\label{Fig:dcj} 
 \end{figure}

  We adopt finite-size scaling $A(g,L)L^{\kappa}=f[(g-g_{ c})L^{1/\nu}]$  to  rescale the binder cumulant  $A=U_2$ or squared magnetization $A={\langle (m^{z})^{2}\rangle}$, in which $\kappa=0$ for $A=U_2$ and $\kappa=2\beta/\nu$ for $A={\langle (m^{z})^{2}\rangle}$, $g=J_{H}(J_{XY}) $ or $ T_E=1/\beta_E$ and $\nu$ is correlation length exponent.
  
  As discussed in the main text, setting $\beta_E=1$ and sweeping $J_{H}(J_{XY})$ near the critical point $J_{c}^P=1$, the data  of $U_2$ can be rescaled by 2D ising universality class (UC ) $\nu=1$ see Fig. \ref{Fig:dcj}. 
  
  However, setting $J_{H}(J_{XY})=1$ and sweeping $T_E=1/\beta_E$  near the critical point $\beta_E=1$, the data  of $U_2$ can not be rescaled by 2D ising UC $\nu=1$ see Fig. \ref{Fig:dcbeta} (a). Set $\nu$ free to fit and we find that the data  of $U_2$ can be rescaled by $1/\nu=0.24(1)$ (see Fig. \ref{Fig:dcbeta} (b)). Furtherly, we set $A={\langle (m^{z})^{2}\rangle}$ and $\kappa=2\beta/\nu$ to fit $\nu$ and $\beta/\nu$. We find that $\beta/\nu=0.123(2)$ and $1/\nu=0.34(2)$ can be used to rescale $\langle (m^{z})^{2}\rangle$ for different system sizes, as shown Fig. \ref{Fig:dcm2}.
 \begin{figure}[htb]
	\centering
	\includegraphics[width=0.46\textwidth]{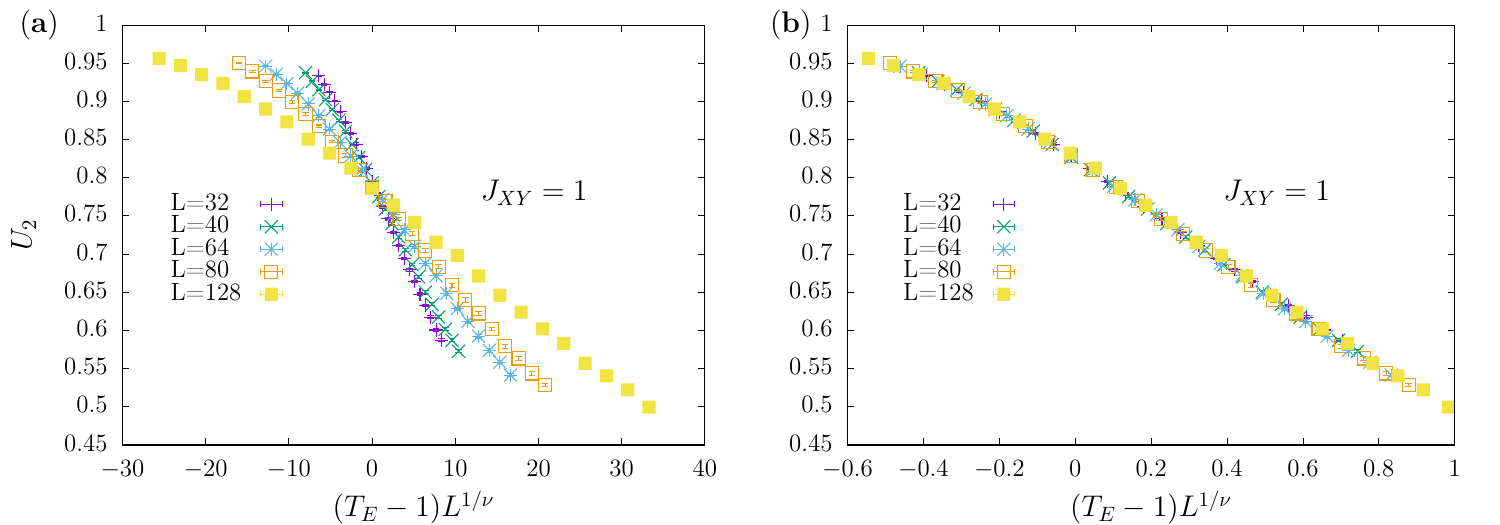}
	\caption{Setting $J_{XY}=1$ and sweeping $T_E=1/\beta_E$ near the critical point $\beta_E=1$, the data  of $U_2$ for different  system sizes can not fall onto the same curve with $\nu=1$ (a) but can fall onto the same curve with $1/\nu=0.24(1)$(b).} 
	\label{Fig:dcbeta} 
 \end{figure}

 \begin{figure}[htb]
	\centering
	\includegraphics[width=0.46\textwidth]{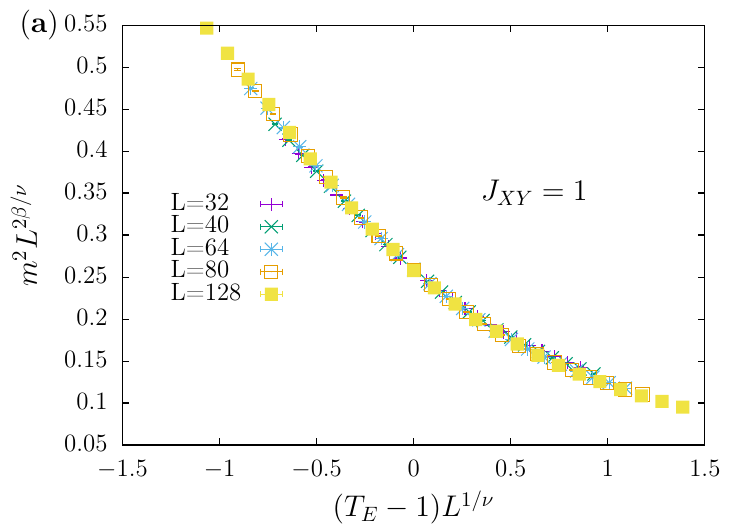}
	\caption{Setting $J_{XY}=1$ and sweeping $T_E=1/\beta_E$ near the critical point $\beta_E=1$, the data  of ${\langle (m^{z})^{2}\rangle}$ for different  system sizes can  fall onto the same curve  with $1/\nu=0.34(2)$ and $\beta/\nu=0.123(2)$.} 
	\label{Fig:dcm2} 
 \end{figure}

\section{The Binder cumulant $U_2$ of the 1D ising model at the finite temperature}
The Binder cumulant $U_2$ of the 1D ising model at the finite temperature is shown in  Fig. \ref{Fig:1dising}.

\begin{figure}[htb]
	\centering
	\includegraphics[width=0.46\textwidth]{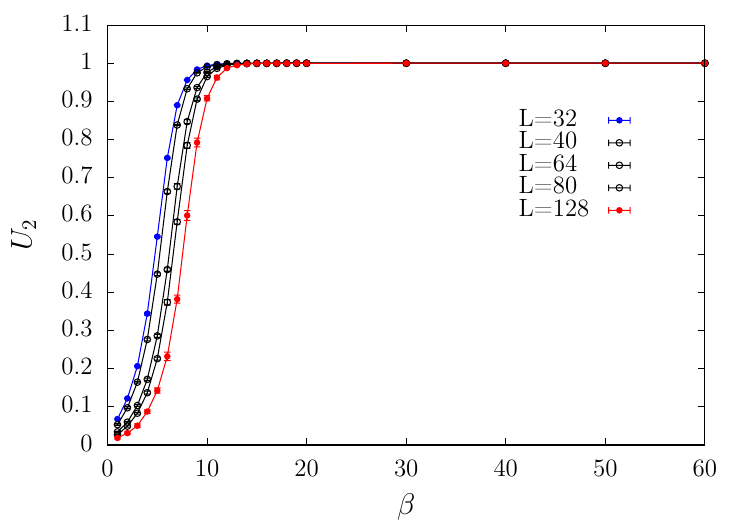}
	\caption{The Binder cumulant $U_2$ of the 1D ising model at the finite temperature } 
	\label{Fig:1dising} 
 \end{figure}

\begin{figure}[htb]
	\centering
	\includegraphics[width=0.46\textwidth]{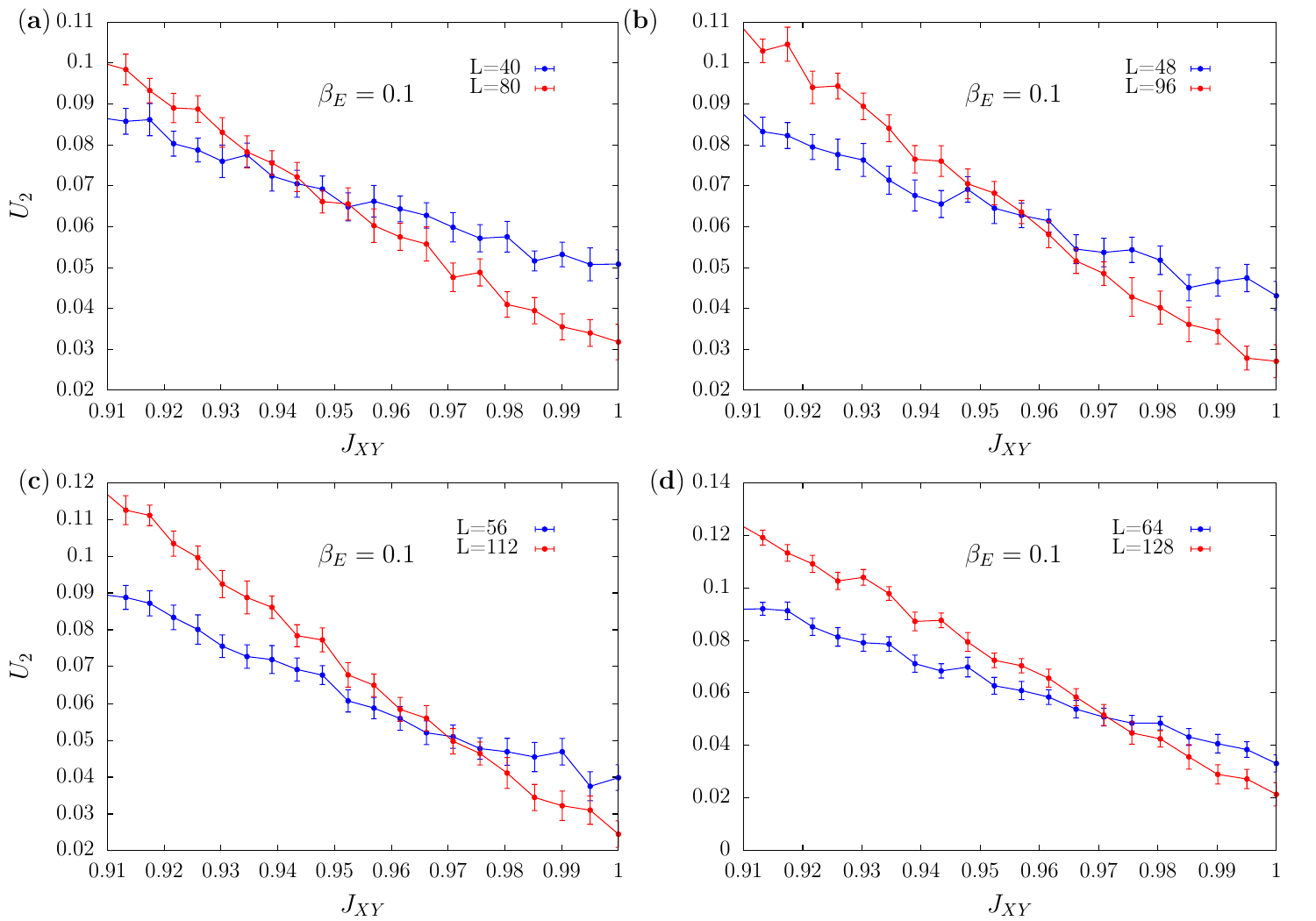}
	\caption{The results of the $U_2$ for different (L, 2L) vs $J_{XY}$ setting $\beta_E=0.1$. } 
	\label{Fig:beta01l2l} 
 \end{figure}

\section{Focus on the OH of the  system $H_{XY}$ in the FM phase}

\begin{figure}[htb]
	\centering
	\includegraphics[width=0.46\textwidth]{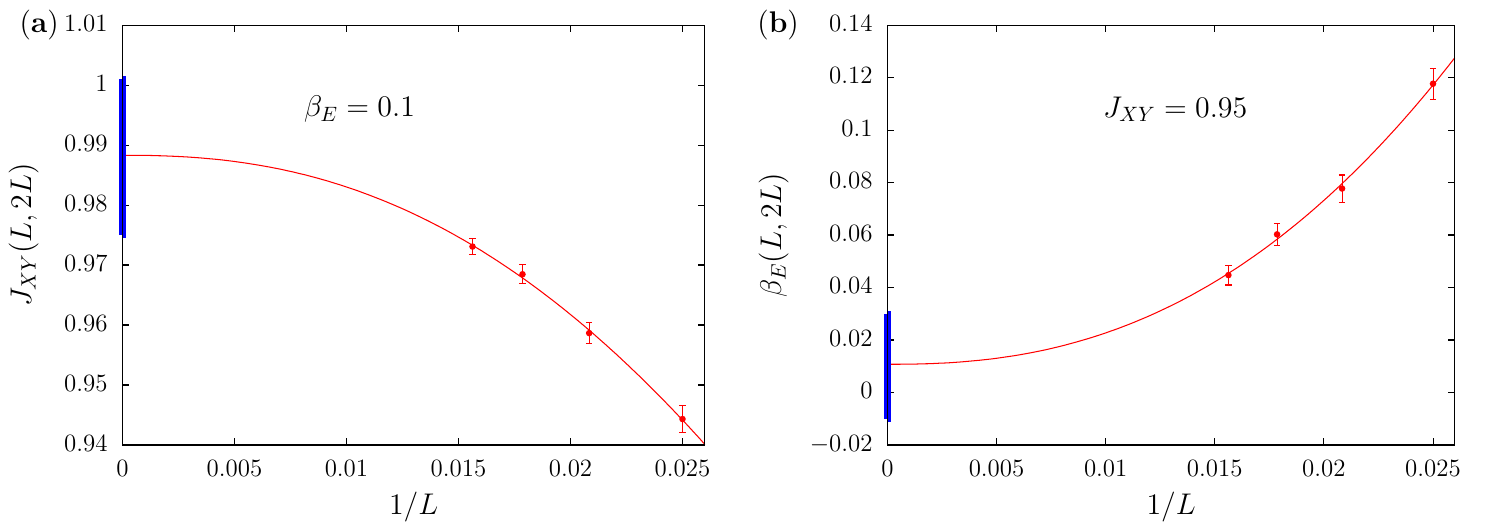}
	\caption{(a) Fitting about  the position of the  crossing points  of the $U_2$  curves setting $\beta_E=0.1$. The blue symbol shows the extrapolated value at the
thermodynamic limit, which is 1 within the error bar. (b) Fitting about the position of the  crossing points  of the $U_2$  curves setting $J_{XY}=0.95$. The blue symbol shows the extrapolated value at the
thermodynamic limit, which is 0 within the error bar.} 
	\label{Fig:orderl2l} 
 \end{figure}

 \begin{figure}[h!]
	\centering
	\includegraphics[width=0.46\textwidth]{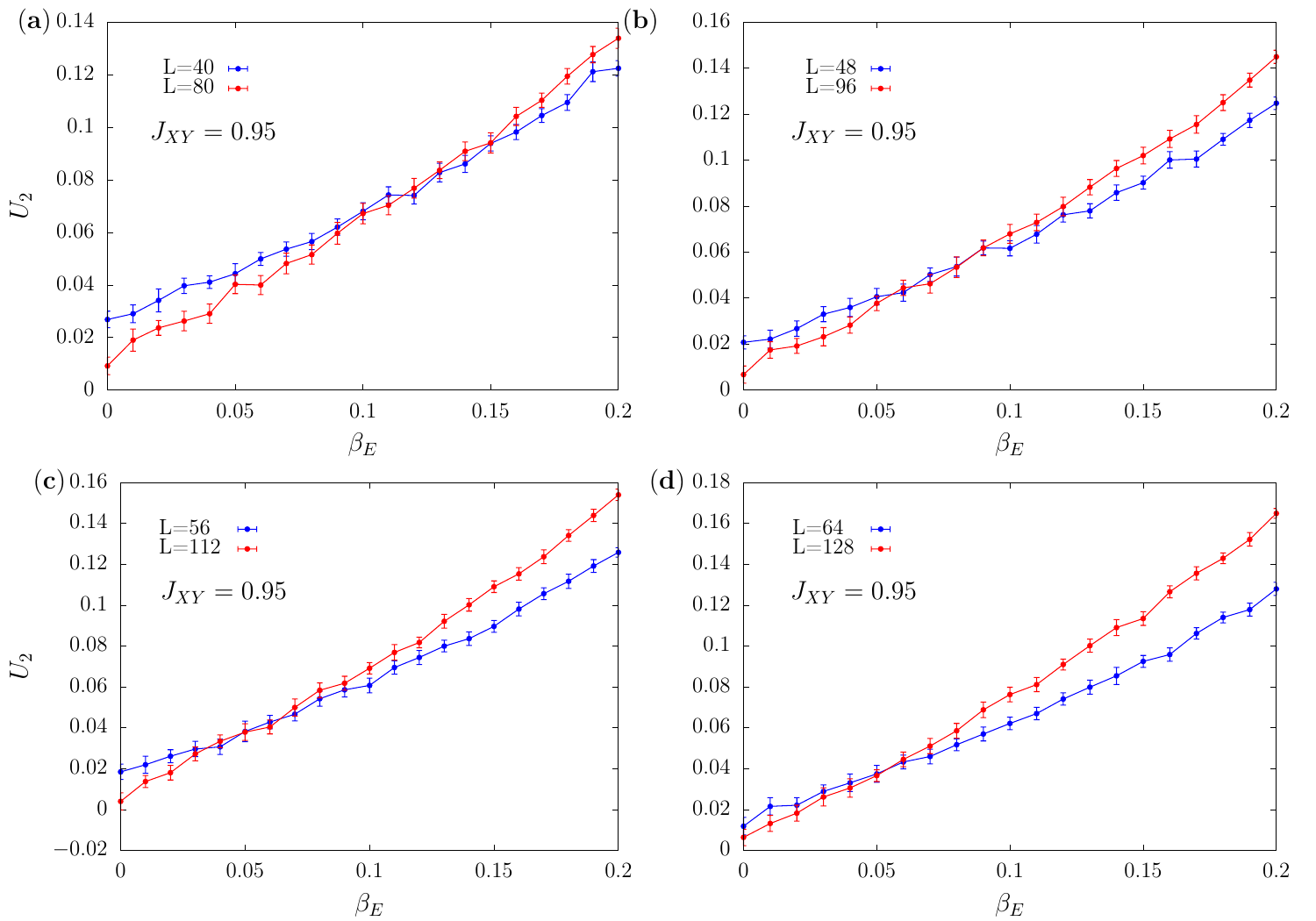}
	\caption{The results of the $U_2$ for different (L, 2L) vs $\beta_E$  setting $J_{XY}=0.95$. } 
	\label{Fig:jxy2.1l2l} 
 \end{figure}
 
As discussed in the main text for binder cumulant $U_2$, the crossings of curves for different sizes drift towards to $J_{XY}=1$ when we set the $\beta_E=0.1$. Further, we show the results of the $U_2$ for different $L$ and $2L$ in Fig. \ref{Fig:beta01l2l} and adopt the standard $(L, 2L)$ crossing analysis to do this.  Fitting the crossing points  of the  curves for $L$ and $2L$ (see Fig. \ref{Fig:orderl2l}) with
\be
J_{XY}(L,2L)= J_{XY}(L\to \infty) +a L^{-\omega},
\ee
we find that $J_{XY}(L\to \infty)$ is 0.988(13) which is 1 within error bar.

Similarly, the results of the $U_2$ setting $J_{XY}=0.95$ for different $L$ and $2L$ are shown  in Fig. \ref{Fig:jxy2.1l2l}. Fitting the crossing points  of the  curves for $L$ and $2L$ (see Fig. \ref{Fig:orderl2l}) with
\be
\beta_E(L,2L)= \beta_E(L\to \infty) +a L^{-\omega},
\ee
we find that $\beta_E(L\to \infty)$ is 0.01(2) which is 0 within error bar.

\section{The details of replica quantum Monte Carlo method}

 \begin{figure}[h!]
	\centering
	\includegraphics[width=0.46\textwidth]{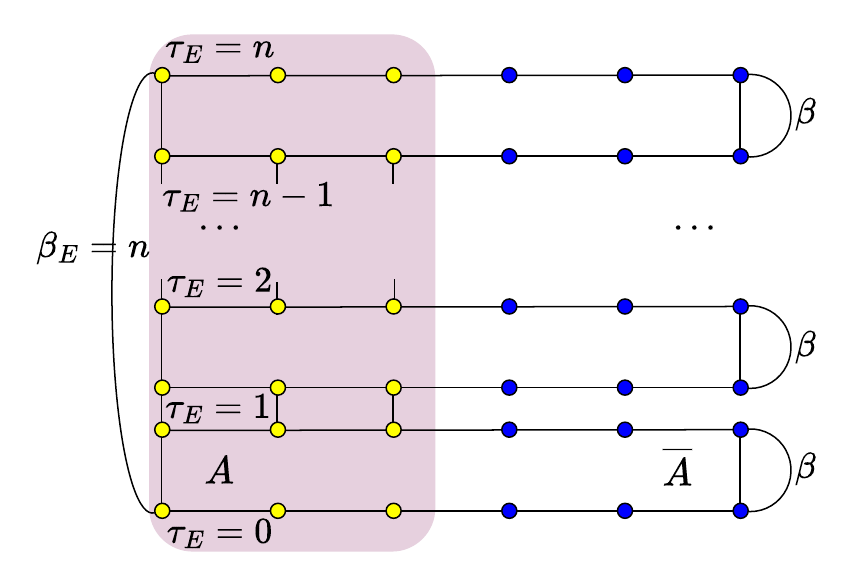}
	\caption{The replica manifold structure in R\'enyi replica-partition-function $Z_A^{(n)}$. The replicas of entangling region A  are glued together in the replica imaginary time direction and each replica of the environment region  $\overline{A}$  is independent in the imaginary time direction. Therefore, the imaginary time length for $H_E$ is $\beta_E=n$  and that for total
system $H$ is $\beta=1/T$. } 
	\label{Fig:replica} 
 \end{figure}

Here, we introduce the details of replica quantum Monte Carlo method. As depicted in Fig.~\ref{Fig:replica}, $Z_A^{(n)}$ is a partition function in a replicated manifold, where the time-boundaries of area $A$ of the $n$ replicas are connected along imaginary time and the time-boundaries of the area $\overline{A}$ of every replica are independent to each other (each replica, the usual periodic boundary condition of $\beta$ is maintained for sites in $\overline{A}$). It can be seen that the effective $\beta_E= n$ for the EH $\mathcal{H}_{E}$ of the subsystem $A$ is in the unit of $1$ whereas the $\beta=1/T$ of the total system is in the inverse unit of the physical energy scale of the original system, $J$ of the Heisenberg model, for instance. 

It's worth noting that there are two concepts of ``temperature'' above, one is the real temperature $T$ ($1/\beta$) for the original Hamiltonian $H$, and the other one is the effective temperature $T_E$ ($1/\beta_E$) for the EH $\mathcal{H}_{E}$. Here we set the $\beta$ as large and let it grow as system size $L$ to make the system approach its ground state. When we talk about the finite temperature phase transition of the EH $\mathcal{H}_{E}$, the ``temperature'' here means the $T_E$ ($1/\beta_E$). In the replica manifold we simulated, the $\beta_E$ can be tuned by changing the number of replicas $n$. 

\end{widetext}
\end{document}